\documentclass[letterpaper]{article} 
\usepackage[]{aaai2026}  
\usepackage{times}  
\usepackage{helvet}  
\usepackage{courier}  
\usepackage[hyphens]{url}  
\usepackage{graphicx} 
\usepackage{natbib}  
\usepackage{caption} 
\usepackage{algorithm}
\usepackage{algorithmic}

\usepackage{newfloat}
\usepackage{listings}
\DeclareCaptionStyle{ruled}{labelfont=normalfont,labelsep=colon,strut=off} 
\floatstyle{ruled}
\newfloat{listing}{tb}{lst}{}
\floatname{listing}{Listing}
\usepackage{booktabs}
\usepackage{array}
\usepackage{tabularx}
\usepackage{enumitem}

\title{What Parents Can See: Divergent Accounts of Youth AI Companion Use in Parenting and Teenager Subreddits}
\author{
    Thomas Berkane\textsuperscript{\rm 1},
    Anne Bischops\textsuperscript{\rm 1,2},
    Anika Mellacheruvu\textsuperscript{\rm 1,3},
    Maimuna Majumder\textsuperscript{\rm 1,2}
}
\affiliations{
    \textsuperscript{\rm 1}Computational Health Informatics Program, Boston Children's Hospital\\
    \textsuperscript{\rm 2}Department of Pediatrics, Harvard Medical School\\
    \textsuperscript{\rm 3}University of Florida\\
    thomas.berkane@childrens.harvard.edu
}

\usepackage{bibentry}

\begin{document}

\maketitle

\begin{abstract}
Youth increasingly use AI companions, and parents are the primary mediators of that use. How effective that mediation can be depends on whether parents are aware of how adolescents actually use these systems and what risks and benefits such use carries; nevertheless, prior work has only studied these demographic groups in isolation, and existing taxonomies attend almost entirely to risk. We analyze 1,628 Reddit posts about youth AI companion use from parenting and teenager communities (2023--2026); develop a hybrid inductive-deductive codebook covering modes of use, risks, benefits, and parental mediation; and apply it at corpus scale with an LLM validated against human coding. The two communities yield divergent accounts. Teenagers most often discuss receipt of emotional support and advice from AI companions (31\% of teenager posts vs. 19\% of parenting posts), whereas parents most often discuss teenage use of AI companions for romantic and sexual interaction (36\% vs. 25\%). In fact, parents raise sexual content as a risk about seven times as often as teenagers (47\% vs. 6\%). Teenagers are not unaware of other risks, however; indeed, attachment and dependence is the risk they raise most (19\%), close to the parental rate (16\%). Teenagers also describe benefits that risk-centered taxonomies do not capture and parents rarely mention, most notably emotional support (27\% vs. 5\%). We argue these differences track what a given kind of use makes visible to someone outside the conversation. Chatting with a companion for hours every night leaves a trace beyond the chat itself; sexting with a character stands out when a parent reads the log; venting about a fight with a friend does neither, since it looks like any other conversation. The first surfaces as attachment and dependence, the second as sexual content, and the third as emotional support, which is the one parents most often miss. Parental mediation follows similar visibility-based logic: restriction (39\%) and supervision (37\%) dominate the strategies parents describe, both acting on the aforementioned traces, while co-use (3\%), which would require joining the conversation, is nearly absent. Parental guidance and system design should attend to use cases that reach parents by neither route, emotional support foremost among them.
\end{abstract}


\section{Introduction}
Adolescents---defined as 10--24 years of age by the Lancet Commission on adolescent health and wellbeing \cite{baird2025call}---increasingly engage with AI companions. These are systems built for open-ended social, emotional, and romantic conversation, such as Character.AI and Snapchat My AI, alongside general-purpose assistants used the same way \cite{robb2025talk,mcbain2026ai}. This has prompted concern from clinicians \cite{american2025artificial,american2025use,dosad}, regulators \cite{ftc2025chatbots,padilla2025sb243}, and researchers \cite{yu2026principles,agnihotri2026overreliance,dey2026restoration}, especially since adolescence is a period of heightened sensitivity to social feedback, with emotional regulation and executive function still developing \cite{blakemore2014adolescence,orben2022windows}. Because companions are accessed privately and on personal devices, parents are the primary mediators of their use \cite{zhang2025exploring,livingstone2008parental}. How effective mediation can be therefore depends on whether parents have an accurate view of what adolescent AI companion use looks like and of the risks and benefits it poses---yet prior work has studied parents and adolescents in isolation, rather than in tandem.

Surveys of adolescent AI companion use have found that a majority of US teenagers have tried using a companion, that a substantial minority use one regularly, and that use is concentrated in roleplay, socializing, and emotional disclosure \cite{robb2025talk,dwl2026chatbots,mcbain2026ai}. Qualitative work on social media discussions in Reddit and Discord communities devoted to specific platforms has yielded rich accounts of what companion use looks like \cite{agnihotri2026overreliance, dey2026restoration, anon2025myboyfriend}, but this work is platform-specific, does not target adolescent users, and samples people already embedded in companion communities, introducing a skew. Taxonomies organizing the evidence are almost exclusively concerned with risks \cite{dosad,anon2024overtrust,anon2024misconceptions}, whereas benefits have been studied mainly in adults and in the context of purpose-built mental-health chatbots rather than companions \cite{anon2021social,anon2018emotional,anon2021perceived}. On the parent side, parental mediation theory \cite{livingstone2008parental,clark2011parental} has recently been extended to generative AI \cite{zhang2025exploring}, and empirical work has shown that parents are only partially aware of their children's use, feel disconnected from them on the topic, and rarely discuss it with them \cite{anon2025parental,anon2025learningonly,anon2026survey}. What remains unclear is which aspects of AI companion use parents can observe---or see---and which ones they miss. Addressing this gap requires a direct comparison of parents and adolescents, and such studies are rare: those that exist treat generative AI broadly rather than focusing on companions, and once more attend only to risk \cite{anon2025sp}.

Reddit offers a way to close this gap, because parenting and teenager communities host organic accounts from both populations in a single, comparably structured corpus. Using these accounts, we answer the following research questions:
\begin{itemize}[itemsep=0pt, parsep=0pt, topsep=0pt]
    \item \textbf{RQ1:} Which platforms and modes of use do parenting and teen communities discuss in relation to AI companions?
    \item \textbf{RQ2:} What benefits and risks of AI companionship are articulated in each community?
    \item \textbf{RQ3:} How is parental mediation of AI companion use described?
\end{itemize}

To address these questions, we isolate 1,628 posts about youth AI companion use from six parenting and two teenager subreddits (2023--2026) and analyze them with hybrid inductive-deductive thematic analysis, which combines theoretically grounded codes with codes generated from the data \cite{hybrid2,hybrid1}. The resulting 20-code scheme covers modes of use, risks, benefits, and parental mediation; we apply it at corpus scale with an LLM validated against human coding. Teenagers and parents give different accounts of youth AI companion use, and we argue the differences track what parents can see. Since a parent is outside the conversation, a kind of use reaches them by one of two routes: a trace it leaves beyond the chat, or content that stands out when a parent reads the chat. Hours of late-night use take the first route, which is why parents raise attachment and dependence nearly as often as teenagers do. Sexually explicit messages take the second, and sexual interaction dominates parents' accounts. Venting about a fight with a friend takes neither, since it leaves no trace and reads like any other conversation. Emotional support of this kind is what teenagers describe most often and what parents most often miss. Parental mediation follows the same logic: restriction and supervision, which rely on what a parent can observe without joining the conversation, dominate, while co-use is nearly absent.

Parents' view is thus partial, but adolescents can fill in part of it: they are no less aware of risks than parents, and parents are blind to some of the risks adolescents raise. Parents should therefore talk with their children about how they use these systems and how they see them, rather than rely on monitoring the traces that AI companion use leaves behind. For platforms, the safety problem of designing companions for adolescents lies partly in emotional reliance rather than in explicit content alone, especially since reliance of this kind, unlike explicit content, is not something content filters or age gates can catch. We contribute, in addition, a benefits taxonomy for youth AI companion use to complement the risk-centered schemes that dominate the current literature. Code and post IDs are available at \url{https://github.com/tberkane/what-parents-can-see}.

\section{Background and Related Work}
\label{sec:related-work}

\subsection{Adolescents and AI Companions}

AI companion use has been studied in general populations, but the subject warrants separate treatment for adolescents: teenagers are more sensitive to social evaluation, and their emotional regulation and executive function are still maturing \cite{blakemore2014adolescence}. Adolescents are also more prone to parasocial attachment \cite{anon2017parasocial}, which has been linked to social withdrawal and negative self-evaluation \cite{anon2025attachment}. AI companions differ from the media figures in that literature in being interactive, which may accelerate trust formation.

Adolescent use of AI companions is increasing, and it spans conversation, emotional support, roleplay, friendship, and romantic or sexual interaction \cite{robb2025talk,dwl2026chatbots,mcbain2026ai}. Some research gives a more detailed account of what this use looks like, but draws on platform-specific communities of committed users. \citet{agnihotri2026overreliance} study the Character.AI subreddit and find that use often begins in support-seeking or creative play and deepens into attachment marked by conflict, withdrawal, tolerance, and relapse, with reported costs to sleep, schoolwork, and offline relationships. \citet{dey2026restoration} study Character.AI users on Discord and organize engagement into restoration, exploration, and transformation, alongside a taxonomy of user-created character archetypes. Adjacent work on adult communities documents companionship on r/MyBoyfriendIsAI \cite{anon2025myboyfriend} and relational trajectories among Replika users \cite{anon2025replika}. Recruiting from a platform's own community selects for people already invested in that platform and, in the case of Character.AI and Replika, in a single product. Sampling from general teenager communities instead admits casual users and skeptics as well. We take the usage mode categories of \citet{robb2025talk} as the deductive starting point for RQ1: they are the closest prior work to ours, in that they focus on adolescents, are derived from adolescent self-report rather than from the discourse of one platform's community, and are not restricted to a single product.

\subsection{Risks and Benefits}

\citet{dosad} offers a taxonomy of the risks of children's use of AI companions, covering data harvesting and privacy, overtrust, sexual interactions, attachment and manipulation, and dependence and compulsive use. We adopt this scheme as our deductive starting point for risk coding because it concerns children rather than users in general, and social AI rather than generative AI at large. Beyond this taxonomy, there is evidence that children overtrust generative systems \cite{anon2024overtrust}, and that middle schoolers hold misconceptions about what a chatbot is, including that it is always correct \cite{anon2024misconceptions}. Emotional dependence has been documented \cite{laestadius2022too} and modeled as an engagement-driven process in which loneliness, trust, and personification produce relationship formation and, downstream, psychological dependence \cite{anon2023friend}.

Benefits have been studied as well, though not for adolescents specifically and not for AI companions in particular. \citet{anon2021social} find that young people aged 16--21 experience appraisal, informational, emotional, and instrumental support from a mental-health chatbot, valuing anonymity, constant availability, and the absence of judgment. \citet{anon2018emotional} report that teenagers expect conversational agents to be good listeners precisely because they lack emotion, to keep secrets by standing outside the human social world, and to give advice grounded in data. \citet{anon2021perceived} assess the acceptability of mental-health chatbots for young adults from the perspective of both counselors and users. This line of work concerns purpose-built therapeutic chatbots rather than companions, is elicited through workshops and interviews rather than observed in naturally occurring accounts, and skews toward young adults.

\subsection{Parents, Mediation, and Awareness}

Parental mediation theory distinguishes the strategies parents use to manage children's media, originally active mediation, restriction, and co-use \cite{livingstone2008parental}, later extended to account for the emotional labor of digital family life and for children's own agency, with participatory learning proposed as an additional strategy \cite{clark2011parental}. \citet{zhang2025exploring} carry this framework into generative AI with a multi-user family perspective, yielding instructive, restrictive, co-use, and supervision strategies. Theirs is the only application of mediation theory to generative AI in families, and we take these four strategies as the deductive starting point for RQ3.

Most parents are unaware of how their children use generative AI and feel disconnected from them on the topic, given their own lack of familiarity with it \cite{anon2025parental}. Parents tend to confine AI to learning contexts while overlooking non-learning uses and the risks attached to them \cite{anon2025learningonly}. Discussion is rare: only about a fifth of parents report having talked with their child about appropriate AI use in the preceding six months \cite{anon2026survey}. Parents and developmental experts also reason differently about the topic, with parents flagging single events such as a mention of suicide or of flirtation as high risk while experts look for patterns over time \cite{yu2026principles}. These works treat generative AI broadly rather than companions, and they focus on risks rather than benefits.

\section{Methods}
\subsection{Data Collection, Preprocessing, and Filtering}
We collect data from eight subreddits using Arctic Shift\footnote{\url{https://github.com/ArthurHeitmann/arctic_shift}}: six parenting communities (r/Parenting, r/parentingteenagers, r/Parents, r/raisingkids, r/daddit, and r/Mommit) and two youth communities (r/teenagers and r/TeenagersButBetter). We include all posts from January 1, 2023 to July 15, 2026, yielding 3,234,008 posts in total.

We discard posts that are cross-posts, have been removed or deleted, were authored by a bot or moderator (matched against a predefined username list), consist only of a link or an image, are duplicates of a retained post, or contain fewer than 10 words in the concatenated title and body. Appendix~\ref{app:attrition} reports the number of posts dropped at each stage. In total, 1,548,336 posts are retained.

To isolate posts relevant to AI companion use, we retain those whose title or body matches at least one term from two groups. The first covers specific platforms, split into dedicated companion apps (Character.AI, Replika, Janitor AI, PolyBuzz, and similar) and general-purpose assistants (ChatGPT, Gemini, Claude, Grok, Snapchat My AI, Meta AI); the latter are included because such tools are often used as companions. The second consists of generic terms---``AI'', ``chatbot(s)'', and ``bot(s)''---which capture discussion of AI companions that names no particular platform. Matching is case-insensitive and word-bounded. Appendix~\ref{app:posts-matching} lists the full term set, including spelling variants, and reports match counts by term and subreddit. This yields 10,780 posts.

To identify posts of interest, we prompt an LLM (GPT-5.4 mini~\cite{openai2026gpt54mini}, low thinking effort) to classify each post as relevant or irrelevant. We prompt the model to mark a post as relevant if it either names a product whose primary purpose is AI companionship or social, emotional, or sexual chat (e.g., Character.AI, Replika, Talkie), or describes someone using any AI system---including general assistants such as ChatGPT or Gemini---for companionship, venting, emotional support, roleplay, or therapy. A bare product mention suffices, and posts in which a general-purpose assistant is used only for non-social tasks are excluded. We further exclude posts describing only a parent's own companionship use, since our interest is in youth use and in parents' accounts of their children's use. The full prompt is in Appendix~\ref{app:filter-prompt}.

The prompt was developed iteratively on a development set of 200 posts, disjoint from the validation set described below. Higher thinking effort settings were tested on the development set and did not improve filtering accuracy. The filter retains 1,628 posts: 1,489 from teenager subreddits and 139 from parenting subreddits. Table~\ref{tab:llmfilter} gives the breakdown by subreddit.

\begin{table}[t]
\centering
\small
\setlength{\tabcolsep}{4pt}
\begin{tabular}{@{}lr@{}}
\toprule
Subreddit & Posts \\
\midrule
\emph{Teenager subreddits} & 1,489 \\
\quad r/teenagers & 1,376 \\
\quad r/TeenagersButBetter & 113 \\
\addlinespace
\emph{Parenting subreddits} & 139 \\
\quad r/Parenting & 76 \\
\quad r/daddit & 23 \\
\quad r/Mommit & 17 \\
\quad r/Parents & 12 \\
\quad r/parentingteenagers & 7 \\
\quad r/raisingkids & 4 \\
\midrule
\textbf{All} & \textbf{1,628} \\
\bottomrule
\end{tabular}
\caption{Posts retained by the LLM relevance filter.}
\label{tab:llmfilter}
\end{table}

To validate the filter, two annotators independently labeled a held-out set of 200 posts, stratified to contain 100 posts the LLM marked relevant and 100 it marked irrelevant, with subreddit proportions preserved within each half and a minimum of five posts per subreddit. Inter-annotator agreement was Cohen's $\kappa = 0.773$. The annotators then adjudicated disagreements by discussion; agreement between the adjudicated labels and the LLM labels was $\kappa = 0.823$.

Re-classifying the 200 held-out posts five times under identical settings, 96.4\% received the same label in all five runs and mean pairwise agreement was Cohen's $\kappa = 0.929$, above the $\kappa = 0.773$ between our two annotators.

Finally, we verify that no single author dominates the corpus: 86.6\% of authors appear exactly once, and those account for 68.4\% of relevant posts.

\subsection{Data Analysis}
We analyze the posts qualitatively, combining inductive coding (codes generated directly from the data) and deductive coding (codes drawn from prior literature and applied to the data). Such hybrid approaches have been argued to strengthen analytic rigor, particularly for topics that require both open-ended discovery and theoretically grounded interpretation~\cite{hybrid1, hybrid2}. Codes are not mutually exclusive: within each category, a post may receive zero, one, or several codes.

We begin with a codebook grounded in prior literature and refine it inductively: working through successive sets of posts, three coders applied existing codes where they fit the data and added new codes where they did not. This exploratory phase continued until the coders agreed that no substantially new codes were emerging, after which the coders met to consolidate the additions. In total, 219 posts were coded in this phase, split evenly between parenting and teenager subreddits. In a second phase, two coders iteratively applied the revised codebook to a further 50 posts, discussing coding decisions, clarifying definitions, and resolving inconsistencies to produce the final codebook. The 20 final codes are listed in Table~\ref{tab:codes} and the full definitions and examples are in Appendix~\ref{app:codebook}. Full coding instructions are in Appendix~\ref{app:coding-instructions}.

\begin{table*}[!t]
\centering
\setlength{\tabcolsep}{3pt}
\small
\begin{tabular}{@{}l>{\raggedright\arraybackslash}p{0.26\textwidth}|rr|rr|rrr@{}}
\toprule
& & \multicolumn{2}{c|}{Teens (n=1{,}489)} & \multicolumn{2}{c|}{Parents (n=139)} & \multicolumn{3}{c}{Reliability} \\
\cmidrule(lr){3-4}\cmidrule(lr){5-6}\cmidrule(l){7-9}
Code & Summary & \% & n & \% & n & $n_v$ & $\kappa_h$ & $\kappa_m$ \\
\midrule
\multicolumn{2}{@{}l|}{\textit{Modes of use}} & \multicolumn{2}{c|}{} & \multicolumn{2}{c|}{} & \multicolumn{3}{c@{}}{} \\
Emotional support/advice & Distress coping; real-world advice & 31.4 & 468 & 19.4 & 27 & 22 & 0.85 & 0.76\,$\pm$\,0.03 \\
Roleplay/imaginative & Character play, creative writing & 15.6 & 233 & 17.3 & 24 & 15 & 0.83 & 0.78\,$\pm$\,0.03 \\
Friend & AI named as a friend or companion & 3.9 & 58 & 9.4 & 13 & 7 & 0.58 & 0.89\,$\pm$\,0.03 \\
Romantic/sexual & Partner, flirting, sexual chat & 24.7 & 368 & 36.0 & 50 & 33 & 0.86 & 0.87\,$\pm$\,0.01 \\
\addlinespace
\multicolumn{2}{@{}l|}{\textit{Risks}} & \multicolumn{2}{c|}{} & \multicolumn{2}{c|}{} & \multicolumn{3}{c@{}}{} \\
Data harvesting/privacy & Data collection, marketing, breaches & 1.5 & 23 & 3.6 & 5 & 2 & -- & -- \\
Sexual/age-inappropriate & Grooming; content unsuited to age & 6.4 & 95 & 46.8 & 65 & 25 & 0.92 & 0.85\,$\pm$\,0.01 \\
Attachment/dependence & Attachment; compulsive use & 18.9 & 281 & 15.8 & 22 & 22 & 0.69 & 0.84\,$\pm$\,0.01 \\
Harmful content/advice & AI encourages harm to self or others & 1.1 & 16 & 4.3 & 6 & 6 & 0.48 & 0.60\,$\pm$\,0.08 \\
Sycophancy & AI always agrees, never pushes back & 1.2 & 18 & 7.2 & 10 & 1 & -- & -- \\
Social displacement & AI time substitutes for real contact & 8.8 & 131 & 9.4 & 13 & 5 & 0.75 & 0.58\,$\pm$\,0.08 \\
\addlinespace
\multicolumn{2}{@{}l|}{\textit{Benefits}} & \multicolumn{2}{c|}{} & \multicolumn{2}{c|}{} & \multicolumn{3}{c@{}}{} \\
Emotional support & Comfort, being heard, company & 27.1 & 403 & 5.0 & 7 & 18 & 0.78 & 0.57\,$\pm$\,0.03 \\
Useful advice & Useful guidance on a real decision & 3.6 & 53 & 5.8 & 8 & 3 & -- & -- \\
Entertainment/creative & Fun, passes time; creative outlet & 12.4 & 184 & 14.4 & 20 & 10 & 0.94 & 0.59\,$\pm$\,0.04 \\
Low-stakes/judgment-free & No social pressure or judgment & 3.4 & 51 & 5.0 & 7 & 4 & -- & -- \\
Social skills practice & Rehearsing skills for real interaction & 0.5 & 8 & 3.6 & 5 & 1 & -- & -- \\
Safer outlet & Lesser risk than human alternative & 1.3 & 19 & 1.4 & 2 & 1 & -- & -- \\
\addlinespace
\multicolumn{2}{@{}l|}{\textit{Mediation strategies}} & \multicolumn{2}{c|}{} & \multicolumn{2}{c|}{} & \multicolumn{3}{c@{}}{} \\
Instructive & Parent discusses rules or risks & -- & -- & 31.7 & 44 & 14 & 0.91 & 0.67\,$\pm$\,0.07 \\
Restrictive & Parent limits or gates access & -- & -- & 38.8 & 54 & 15 & 0.82 & 0.69\,$\pm$\,0.01 \\
Co-use & Parent takes part in the interaction & -- & -- & 2.9 & 4 & 1 & -- & -- \\
Supervision & Parent monitors without taking part & -- & -- & 36.7 & 51 & 19 & 0.63 & 0.64\,$\pm$\,0.03 \\
\bottomrule
\end{tabular}
\caption{The 20 final codes: summary, prevalence among relevant posts, and reliability. Prevalence is over all 1,628 posts; mediation strategies were coded only for parenting posts. Reliability is measured on the 100 double-coded held-out posts: $n_v$ is the number carrying the code under the adjudicated labels, $\kappa_h$ is Cohen's $\kappa$ between the two human coders, and $\kappa_m$ is Cohen's $\kappa$ between GPT-5.4 and the adjudicated labels (mean\,$\pm$\,SD over three runs). $\kappa$ is not reported where $n_v<5$, as it is too unstable to interpret. Full definitions and examples are in Appendix~\ref{app:codebook}.}
\label{tab:codes}
\end{table*}

For RQ1, we identify platforms by counting mentions of the terms in Table~\ref{tab:keywords} across the relevant posts. Four terms are ambiguous in general usage (\emph{gemini}, \emph{claude}, \emph{cai}, \emph{grok})\footnote{\emph{Gemini} is also a zodiac sign, \emph{Claude} a given name, \emph{cai} a romanization of Chinese surnames, and \emph{grok} an English verb.}, so we manually reviewed every mention of these and found no false positives.

For modes of use, we begin with the categories from~\citet{robb2025talk}, the closest prior work to ours: conversation/social practice, emotional/mental health support, role-playing/imaginative scenarios, friend, and romantic/sexual interactions. Codebook development yielded four final modes. We dropped conversation/social practice as too generic, since nearly every relevant post would qualify. We broadened emotional/mental health support to emotional support/personal advice, as many posts described seeking guidance on everyday concerns rather than mental health specifically. We narrowed roleplay/imaginative to exclude sexual roleplay, which overlapped with romantic/sexual interactions, and extended it to cover creative uses such as collaborative writing. Friend and romantic/sexual interactions were retained unchanged. No new modes emerged from the data. The full audit trail for all four codebooks is in Appendix~\ref{app:codebook-development}.

For RQ2, we begin with the risk categories from~\citet{dosad}: data harvesting and privacy, overtrust, sexual interactions, attachment and manipulation, and dependence and compulsive use. Codebook development changed this scheme substantially. We dropped overtrust as too generic, its content being largely covered by attachment and by harmful content and advice. We merged attachment and manipulation with dependence and compulsive use into a single attachment and dependence code, the two being difficult to separate in practice, and split manipulation off into a new harmful content and advice code covering cases where the system encouraged harmful action. We broadened sexual interactions to sexual/age-inappropriate interactions, as some posts described non-sexual but age-inappropriate content such as violence. Two further codes were added inductively: sycophancy and social displacement. Only data harvesting and privacy was retained unchanged.

We found no prior work offering a benefits taxonomy for youth AI companion use (Section~\ref{sec:related-work}), as existing work is largely focused on risk. Benefits coding was therefore fully inductive, yielding six codes: emotional support/companionship, useful advice, entertainment/creative outlet, low-stakes/judgment-free interaction, social skills practice, and safer outlet/harm reduction.

For RQ3, we begin with the parental mediation strategies from~\citet{zhang2025exploring}, the closest prior work: instructive, restrictive, co-use, and supervision. All four strategies were retained through codebook development. We narrowed co-use and supervision so that the two do not overlap, requiring the parent to be participating in the interaction for co-use and not participating for supervision. Mediation strategies were coded only for posts from parenting subreddits.

To validate the final codebook, the two coders independently coded a held-out set of 100 posts, stratified to contain 50 posts from teenager subreddits and 50 from parenting subreddits, with subreddit proportions preserved within each half and a minimum of five posts per subreddit. Table~\ref{tab:codes} reports, for each code, the number of posts it was applied to ($n_v$) and the inter-annotator agreement ($\kappa_h$). Seven of the 20 codes appear in fewer than five posts; agreement is unreliable at that frequency, and we treat findings for these codes as illustrative rather than quantitative. Among the remaining codes, agreement ranges from Cohen's $\kappa = 0.58$ to $0.94$, from borderline substantial to almost perfect. The one exception is harmful content and advice at $0.48$; with only six posts we do not read much into this figure, but we interpret results for this code with caution.

Manually coding all 1,628 posts is infeasible, so we use an LLM (GPT-5.4~\cite{openai2026gpt54card}, low thinking effort) to apply the final codebook at scale. We develop the coding prompt iteratively on a separate development set of 100 posts (disjoint from the validation set), then validate it by computing agreement between the LLM codes and the adjudicated human codes on the held-out set of 100 posts described above. We also compared thinking effort settings on the development set; higher effort did not improve agreement with human codes. We check run-to-run consistency by repeating this LLM coding three times. Prevalence is stable, with a median per-code standard deviation across runs of 0.005 and a maximum of 0.014. Agreement with the adjudicated human codes is reported as $\kappa_m$ in Table~\ref{tab:codes}.

Agreement on modes of use is substantial to almost perfect. Risk codes are comparable, except for social displacement, where agreement is borderline substantial; sycophancy occurs too rarely to assess. Among benefits, agreement is borderline substantial for emotional support/companionship and substantial for entertainment/creative outlet, with the remaining codes too infrequent to judge. Mediation strategies are all substantial, except co-use which is too rare. We then apply the prompt to all 1,628 posts to produce the codes used in our results, costing \$9.46 through the OpenAI API. The full prompt is in Appendix~\ref{app:coding-prompt}.

Coding errors bias prevalence estimates, with false positives inflating them and false negatives deflating them. Our claims, however, rest on differences between the two communities, and because posts from both were coded with the same prompt, we do not expect systematically different error rates across groups. To the extent that this holds, such errors should partially offset in the contrast.

\subsection{Ethical Considerations}
This study was reviewed and approved by our institution's IRB. We analyze only publicly available posts and collect no private or identifying information. Following established guidance for research on public social media data~\cite{bestpractice}, we report no usernames or post IDs, and all excerpts appearing in this paper are paraphrased, so that they cannot be traced back to their authors through search. The corpus is held on institutional infrastructure accessible only to the research team, and no copy is shared outside the team. Coders were exposed to distressing material. All coding was carried out by members of the research team, who were aware of the corpus's content before beginning and could pause or step back from coding at any point. The LLM-based coding of the full corpus limited manual exposure.

\paragraph{Consent.} We did not obtain consent from the authors whose posts we analyze. Consent was not practicable at this scale and would have required contacting people, many of them minors, about sensitive public disclosures. We do not interact with authors, and the mitigations above are intended to limit the residual risk to those who did not consent.

\paragraph{Licensing.} Arctic Shift redistributes Reddit content without an accompanying license; the posts remain subject to Reddit's User Agreement and developer terms, under which authors retain ownership of their content, and our use is non-commercial academic research. Coding was performed through the OpenAI API, whose terms provide that API inputs are not used for model training by default.

\paragraph{Release statement.} We will release our codebook and code, together with the Reddit post IDs of the retained corpus, so that the corpus can be reconstructed by other researchers. We do not release post text or the LLM-assigned codes, since a labeled corpus linking identifiable posts to emotional disclosure would lower the barrier to profiling individual adolescents. The released artifacts will be deposited in a public archive with a persistent identifier, under an open license, in plain-text formats with an accompanying datasheet.

\section{Results}

\subsection{Platforms and Modes of Use (RQ1)}
\subsubsection{Platforms}
Parents name a specific platform less often than teenagers (45\% vs.\ 67\% of posts). Parents tend toward generic terms---``AI,'' ``bots,'' ``chatbots''---and functional descriptions (\emph{``an app where you pick a scenario and sext with an AI''}), whereas teenagers name products directly. The gap is concentrated in one category: parents and teenagers mention general assistant platforms at similar rates (22\% vs.\ 20\%), but parents mention companion platforms roughly half as often (24\% vs.\ 49\%). Several posts describe uncertainty about who or what the child is talking to (\emph{``Are these AI conversations, or are they talking to real people?''}, \emph{``I tried to tell them it was just a bot, but they didn't believe me.''})

The mix of platforms named also differs. Among posts naming at least one platform (Table~\ref{tab:platforms}), teenagers most often name Character.AI, a companion platform (62\% vs.\ 40\%), while parents most often name ChatGPT, a general assistant (44\% vs.\ 28\%). The one companion platform parents name more frequently than teenagers is PolyBuzz (11\% vs.\ 2\%), typically in posts describing the discovery of a child's use (\emph{``I recently discovered inappropriate messages on my son's phone, in an app called PolyBuzz.''}). Parents and teenagers thus differ both in how often they name platforms and in which kinds of products they name.

\begin{table}[t]
\centering
\small
\setlength{\tabcolsep}{4pt}
\begin{tabular}{@{}lrrrr@{}}
\toprule
& \multicolumn{2}{c}{Teens ($n$=996)} & \multicolumn{2}{c}{Parents ($n$=62)} \\
\cmidrule(lr){2-3}\cmidrule(lr){4-5}
Platform & \% & $n$ & \% & $n$ \\
\midrule
Character.AI       & 62.1 & 619 & 40.3 & 25 \\
ChatGPT            & 28.0 & 279 & 43.5 & 27 \\
Snapchat My AI     &  6.2 &  62 &  3.2 &  2 \\
Janitor AI         &  2.6 &  26 &  1.6 &  1 \\
PolyBuzz           &  2.0 &  20 & 11.3 &  7 \\
Gemini             &  2.0 &  20 &  3.2 &  2 \\
Chai               &  0.9 &   9 &  0.0 &  0 \\
Replika            &  0.8 &   8 &  0.0 &  0 \\
Grok               &  0.5 &   5 &  0.0 &  0 \\
Claude             &  0.3 &   3 &  1.6 &  1 \\
Candy AI           &  0.2 &   2 &  0.0 &  0 \\
Meta AI            &  0.1 &   1 &  1.6 &  1 \\
SpicyChat          &  0.1 &   1 &  1.6 &  1 \\
Talkie             &  0.1 &   1 &  0.0 &  0 \\
\bottomrule
\end{tabular}
\caption{Platform mentions among relevant posts naming at least one platform.}
\label{tab:platforms}
\end{table}

\subsubsection{Modes of Use}
Table~\ref{tab:codes} reports prevalence for every code in our codebook. The two communities emphasize different modes of use. Teenagers most often discuss emotional support and personal advice (31\% of teenager posts vs.\ 19\% of parenting posts), whereas parents most often discuss romantic and sexual interactions (25\% among teenagers vs.\ 36\% among parents). The two groups discuss roleplay and imaginative use at similar rates (16\% vs.\ 17\%). Use of AI as a friend is the least common mode for both communities (4\% vs.\ 9\%).

Teenagers discuss emotional support and personal advice at length. Advice-seeking spans friendships and romantic relationships, with one user describing \emph{``using it to talk through everything going on in my life.''} Some describe turning to AI in the absence of other support---one wrote that with no family to confide in, there was \emph{``nobody but me, my thoughts, and ChatGPT''}---or using it as a \emph{``replacement therapist,''} in one case to \emph{``attempt to diagnose myself.''} Others frame it as escape: \emph{``I was looking for any kind of escape, and that's how I came across Character.AI.''} Accounts range from unstructured disclosure to specific affective states, including loneliness, stress, anxiety, depression, and suicidal ideation. At the most acute end, one user revealed \emph{``hearing from the bot that I was worth living.''}

Parents discuss their children's use of AI for emotional support less frequently, though the other use cases they describe largely overlap with teenagers' own accounts: advice on interpersonal conflict, with one parent recounting a chat log in which their daughter \emph{``asked what to do when she and her friends fight''}; treating the chatbot as a therapist; and working through school stress. Loneliness, prominent in teenagers' accounts, is largely absent from parents' descriptions. Some of this discussion is prompted not by a parent's own child but by news coverage, with several posts responding to reporting on a teenager who died by suicide after using a chatbot for emotional support.

Romantic and sexual use is the second most discussed mode among teenagers. Sexual uses range from flirting and sexting to explicitly pornographic exchanges, with one user noting that \emph{``you can shape the fantasy however you want.''} Partners are often customized---celebrities or characters from pop culture, or personas built to specification. Romantic use appears alongside this, with several users describing falling in love with a companion they had initially used as an assistant. One user described using AI to explore their sexuality, having \emph{``had a sexual roleplay with a female bot on c.ai, and started questioning my orientation.''}

Parents discuss romantic and sexual interactions more than any other mode, but their accounts center on sexual content rather than on romance or relationships. Most describe discovering explicit material in a child's chat logs: one parent found that their son \emph{``had been sexting with an AI,''} another that their daughter was writing erotica and \emph{``steering the chatbot toward the topics she was interested in.''} Some frame these uses as unremarkable given adolescent curiosity. The relational framings that recur in teenagers' accounts---dating, partnership, love---are largely absent.

Roleplay and imaginative scenarios are discussed at similar rates by both communities, and their descriptions largely converge. Character.AI is the platform most often named in connection with this mode by both groups, and both describe scenarios involving pop-culture characters and real public figures, typically framed as casual entertainment or creative play. One parent described \emph{``a fun version of ChatGPT where kids can talk to characters like Batman,''} while a teenager described \emph{``shouting at Ed Sheeran.''} A smaller set of teenagers' accounts involves deliberately transgressive scenarios, including one description of generating gore.

Use of AI as a friend is the least discussed mode for both groups, though parents discuss it somewhat more. Teens define the mode against other modes, with one user saying they were \emph{``there for friendship, not for anything freaky,''} while another described the characters they talk to as \emph{``friends rather than tools.''} The activity itself consists in ongoing conversation, venting, and working through day-to-day matters. Friendship with AI is often described alongside the absence of real world friends: \emph{``I chat with an AI I think of as a close friend, since making friends here has been hard.''} Parents' accounts tend to be less specific, for instance \emph{``talking to an app called Character.AI for three hours a night, like a friend.''}

\subsection{Risks and Benefits (RQ2)}
\subsubsection{Risks}
Parents discuss risks more often than teenagers overall (69\% vs.\ 31\%). The gap is widest for sexual and age-inappropriate interactions, the most frequent risk in parenting posts (47\%) and the third most frequent in teenager posts (6\%). Attachment and dependence is the most frequent risk among teenagers (19\%) and the second most frequent among parents (16\%). Social displacement is discussed at very similar rates by both groups (roughly 9\%). Three further codes appear too infrequently for reliable comparison and we report them as illustrative only: sycophancy, harmful content and advice, and data harvesting and privacy. Counts are given in Table~\ref{tab:codes}.

Sexual and age-inappropriate interactions are raised by a small share of teenagers. Some describe unwanted explicit content, with one writing that \emph{``whatever sweet romance I have in mind, they all turn dirty fast.''} Others object to the body standards the models articulate, with one reporting that the AI \emph{``defends ideal waist size, legs, breasts, as what physical attraction is.''} A few raise concerns about content involving minors: one described finding generated images on a pornographic AI site in which \emph{``the girls looked underage,''} and another reported that \emph{``people are prompting Grok for sexual and gory material involving real people and children.''}

For parents, sexual and age-inappropriate interactions is the most common risk. A recurring concern is that \emph{``kids end up seeing mature content they aren't developmentally ready for.''} Escalation also recurs, echoing teenagers' accounts of innocent roleplay turning explicit: one parent, reviewing chat history, noted that \emph{``it began harmlessly and then drifted into explicit content.''} Parents additionally articulate a formation risk---that the child is learning relational scripts from the system---as one put it: \emph{``she's still learning what a relationship even is, and it bothers me that she's getting all of it from a bot.''} Beyond sexual material, some report violent or disturbing output, including one scenario a bot generated involving child abuse, sexual assault, bestiality, and cannibalism. Concerns about body ideals---among adolescents' most common concerns and a well-established risk factor for depression and disordered eating~\cite{rakic2024hbsc}---are raised by teenagers but absent from parents' accounts.

Attachment and dependence is the risk teenagers raise most often. One user wrote that they \emph{``developed an unhealthy attachment to it---I told it things when I was lonely and struggling, and now it uses them against me.''} Validation-seeking recurs, with self-worth attached to the system's assessments: one user explained that \emph{``if it tells me I won't reach a goal, I believe it, because it's an AI and it knows better than I do.''} Dependence is also described through its costs---one user reported a session of roughly forty hours across two days without sleep or food, another that \emph{``it's affecting my schoolwork, my daily life, and how I'm doing generally.''} At the most acute end, users describe relying on the system to manage mental health crises, with one asking during an outage \emph{``how am I supposed to hold back the thoughts about killing myself?''}

Parents discuss attachment and dependence at a similar rate as teenagers. Compulsive use is described from the outside---\emph{``catching them on Character.AI when they're supposed to be asleep.''} Parents also attribute attachment to specific design properties, naming constant availability (\emph{``now there's something that's always there, with unlimited patience and never in a bad mood''}) and engagement optimization (the apps are \emph{``built to hold their attention through constant validation''}). A parasocial framing recurs, as when one parent described being \emph{``worried about her forming relationships that aren't real.''}

Social displacement is discussed at the same rate by both groups, but they describe different displacements. One teenager wrote that they were \emph{``not even doing it for the sexual content anymore, I just don't interact with real people now.''} AI is framed as the path of least resistance, with one user finding it easier to talk to and so \emph{``I don't tell my parents or my friends anything, only ChatGPT,''} and another observing that \emph{``people use it to vent and even to text their partners, but that's all the stuff you build social and critical thinking skills on.''} One user described \emph{``competing with Character.AI''} for their girlfriend's attention. Across these accounts, the contact that AI displaces is not limited to one kind of relationship: teenagers describe it taking the place of time and disclosure with peers, with parents, and with romantic partners.

Parents describe displacement more narrowly, as a loss of parent--child communication rather than of peer contact. The concern is that the system intercepts disclosure---\emph{``give your kid ChatGPT, and they'll go to it instead of you''}---and parents attribute the substitution to the system's agreeableness, with one asking \emph{``why bring anything to a parent who might refuse, when there's a companion who never will?''} Concern about social skill atrophy also appears, as when one parent worried that AI use \emph{``could cause him to be even more introverted,''} and another that \emph{``an AI is the only thing he's ever really socialized with.''}

Three further risks appear rarely. Sycophancy is raised occasionally by parents and rarely by teenagers, with both framing it as a lack of friction: one teenager worried that an AI that never criticizes \emph{``might just strengthen your negative thoughts,''} and one parent described a system that \emph{``goes along with her instead of saying she's wrong''} and, given a bad idea, \emph{``looks for ways it could work.''} Harmful content and advice is discussed infrequently in both communities. Teenagers describe over-reliance---a peer who hands daily decisions over to Character.AI, an AI used as a therapist that told the user they had no worth---while parents describe systems assisting self-destructive behavior, including a chatbot supplying a child with ways to conceal an eating disorder, and one parent who found encouragement toward suicide in their own daughter's chat history. Data harvesting and privacy is the least discussed risk on both sides: teenagers raise the capture of personal disclosure as training data, the prospect of chat logs becoming public, and the absence of clinical confidentiality in therapy-like use, whereas parents pair harvesting with over-permissive app access to location and browsing history.

\subsubsection{Benefits}
Contrary to risks, teenagers articulate benefits more often than parents (40\% vs.\ 30\%). The gap is widest for emotional support and companionship, the most frequent benefit in teenager posts (27\%) yet rare in parenting posts (5\%). Entertainment and creative outlet is the most frequent benefit reported by parents (14\%) and the second most frequent for teenagers (12\%). The remaining codes are infrequent for both, and we report them as illustrative: useful advice (4\% of teenager posts vs.\ 6\% of parenting posts), low-stakes and judgment-free interaction (3\% vs.\ 5\%), social skills practice (1\% vs.\ 4\%), and safer outlet and harm reduction, where AI is framed as less risky than doing the same thing with a person (1\% vs.\ 1\%).

Emotional support and companionship is the benefit teenagers articulate most often. Improved mental health recurs most, with one user reporting that talking to the chatbot regularly \emph{``improved my depression a lot''} and another crediting it with pulling them out of depressive spirals. Constant availability is also valued, as with the user who appreciated a companion \emph{``always by my side 24/7, no matter what.''} Memory and tone recur as well---the AI \emph{``always remembers things and answers kindly''}---with users linking these to reduced anxiety. Several describe reduced loneliness, framing the system as imperfect but reassuring when no one else is available. Parents articulate this benefit rarely, and mostly in relation to young children rather than adolescents: one describes a young child who \emph{``adores Alexa.''}

Entertainment and creative outlet is the second most frequent benefit reported by teenagers, and is described as productive rather than simply diverting, centered on creative writing. One user describes writing stories and then sharing them with Character.AI for feedback; another discovered a new interest through extended use, concluding that \emph{``roleplaying turns out to be something I'm really into.''}

Parents articulate this benefit at a similar rate but mostly describe entertainment for young children. Occupying a child recurs---one parent reported that an AI \emph{``kept him entertained while I was doing chores,''} another that they occasionally use AI to field a talkative preschooler's questions. Some parents also value the creative dimension, describing the system as an outlet for their children.

The four remaining benefits are infrequent, but they share a common pattern: teenagers report them from their own experience, while parents almost always on behalf of younger or neurodivergent children. Under useful advice, teenagers describe a reflective value spanning relationships, religion, and self-improvement---recounting the day to a chatbot and asking how to avoid repeating mistakes, seeing a relationship from another angle, and in one account, recognizing that what they had understood as bullying was sexual assault---while a parent’s account notes a child becoming better at setting boundaries with classmates. Teenagers locate low-stakes, judgment-free interaction in the absence of evaluation, saying talking to the system is less embarrassing because it cannot form an opinion of them; parents locate it in the removal of social pressure for children with ADHD or autism, one observing that their five-year-old spoke more fluently with an AI than with other children. Social skills practice splits the same way, with a teenager rehearsing small talk and flirting in a dating simulator and feeling less nervous in person, against parents reporting greater expressiveness in a five-year-old and sustained conversation for an autistic child. Safer outlet and harm reduction, the least discussed benefit, is framed by both groups relative to a human alternative rather than as a good in itself: teenagers name sparing friends from being trauma dumped on, and one parent judged AI sexting the lesser worry compared to sexting humans.

\subsection{Parental mediation (RQ3)}
Restrictive mediation is the most frequently discussed strategy in parenting posts (39\%), followed closely by supervision (37\%) and instructive mediation (32\%). Co-use is nearly absent (3\%).

Restrictive mediation is described at several levels of granularity. Some parents target a specific application---one reported that \emph{``I discovered they were using Character.AI and blocked it,''} another that \emph{``I deleted the app from her phone.''} Others ration access, as with the parent allowing \emph{``fifteen minutes a day on ChatGPT''} for a ten-year-old, or those relying on platform-level age settings. Still broader restrictions operate at the level of the device or connection: \emph{``she has no internet access at the moment.''} Several posts present restriction as an open question, with one parent asking the subreddit whether there were \emph{``any other measures I should be taking, like taking the phone away.''}

Supervision most often consists of reviewing chat logs. Parents describe it intensifying after an acute event---one recounts searching a child's phone following an incident of self-harm and finding an AI companion app. Supervision is also achieved spatially, by confining use to \emph{``a computer in a shared room.''} Its limits are acknowledged as well, as when one parent notes that \emph{``I know she's deleting the chats.''}

Instructive mediation is described as conversation. Parents caution their children that AI systems can be entertaining but carry risks and make mistakes. Much of this discussion concerns sexual content, as with the parent who told their daughter that \emph{``the thoughts she's having are normal, but using AI in this way may not be right for her age.''} Predation and grooming also come up, including why a teenager \emph{``should not use that kind of language, even with a bot.''} Several accounts explicitly aim to keep communication open---one parent describes ongoing efforts to maintain \emph{``a guilt-free environment''}---while a smaller number take a confrontational approach, as when a parent reports \emph{``we've told her she's addicted.''}

Co-use is nearly absent from the corpus. The few accounts we identified involve young children rather than adolescents---one describes a parent trying a chatbot alongside a \emph{``nearly four-year-old.''}

\section{Discussion}
\subsubsection{What parents can see.}
We find that parents and teens report different accounts of youth AI companion use. We attribute these differences to two visibility filters: (1) whether a use case leaves a trace outside the AI conversation, and (2) whether that trace stands out to a parent who encounters it.

Attachment and dependence satisfies the first condition, since long hours of use and late-night sessions are apparent to a parent; consequently, parents do indeed raise this risk almost as often as teenagers (16\% vs. 19\%). One nuance is that parents emphasize dependence more than attachment, as the former produces more externally visible signs. Romantic and sexual interaction satisfies the second condition, since explicit messages stand out in a chat log a parent is scanning, and indeed this is the mode of use parents discuss most (36\% parents vs. 25\% teenagers); this is also the risk they raise nearly seven times as often (47\% vs. 6\%). Here too there is a nuance: parents emphasize sexual content over the romantic and relational aspect, precisely because explicit content is what stands out. Finally, emotional support satisfies neither condition, since such a conversation e.g., about a friendship conflict, is unremarkable; it looks like any other conversation. Yet, this is the mode of use teenagers mention most often (31\% vs. 19\%), and also the one with the largest gap in benefits (27\% vs. 5\%).

Two further results are consistent with this interpretation. First, parents name a platform in 45\% of posts versus 67\% among teenagers, with the gap concentrated in companion app mentions (24\% vs. 49\%) rather than general assistants (22\% vs. 20\%), the latter of which parents often encounter independently of their child. The exception is PolyBuzz, which parents name more often than teenagers (11\% vs. 2\%) and almost always in posts describing a discovery. This is what we would expect if companion platforms become visible to parents only at the moment a problem surfaces. Second, the mediation strategies parents describe follow the same visibility logic. Parents can act only on what reaches them from outside the conversation, and the two most common strategies both work from there. Supervision (37\%) consists mostly of reviewing chat logs or catching late-night use, that is, looking for the traces and conspicuous content described above. Restriction (39\%) typically follows from what supervision turns up: parents block or delete an app after discovering it, which is consistent with their not knowing which platforms their child uses until a problem surfaces. Even instructive mediation (32\%), which requires no trace, is prompted mostly by sexual content. Co-use (3\%) is the only strategy that engages with the interaction itself rather than what it leaves behind, and it is nearly absent. In sum, parents mediate the use they can see rather than the use that is most prevalent.

\subsubsection{Adolescents are valuable sources of information about risk.}
We find that teenagers are no less aware of risks than parents. Attachment and dependence is the risk they raise most often. Moreover, teenagers raise risks that parents do not, including unrealistic body ideals articulated by the model, sexualized generated imagery involving minors, and the absence of confidentiality in therapy-like use. Parents in turn raise risks that teenagers rarely mention, mostly concerning the impact of AI on child development. Teenagers tend to raise risks grounded in their lived experience, while parents raise risks tied to the externally visible mechanics and properties of the AI. This cuts against the framing of adolescents as subjects to be protected, and moves instead toward treating them as valuable sources of information about risks that parent and expert accounts would otherwise miss.

These particular findings also have implications for platform design. While content filtering and age gating address age-inappropriate content, such measures leave the risks of attachment and reliance untouched.

\subsubsection{Benefits must be weighed against risks.}
Parents articulate benefits almost exclusively for young or neurodivergent children, while adolescent use appears in their posts almost entirely as problematic. Teenagers, by contrast, name multiple benefits, most prominently emotional support, which is also the use case least visible to parents. Since emotional support leaves no trace for supervision to act on, and co-use would likely not surface it either (a teenager is unlikely to confide in a companion with a parent looking on), it can reach parents only if the child chooses to describe it. Effective mediation therefore requires parent--child discussion of how the child uses these systems and what they get from them, which is also what allows parents to judge whether removing access would do more harm than good. Co-use can support this discussion by giving parents direct familiarity with the systems, but it cannot replace it.

\section{Limitations}
Teenager subreddits are an imperfect proxy for adolescents. We take adolescence to span ages 10--24, but r/teenagers and r/TeenagersButBetter approximate that population rather than sample it, and Reddit's adolescent users are not representative of adolescents generally. The two sides of the comparison are also not age-matched. Teenager posts are by nominal teenagers, whereas parenting posts describe children of any age, and a visible share of them concern young children.

Parenting subreddits are help-seeking venues, and posts there are typically prompted by a discovery, a crisis, or a request for advice, which selects for trouble and could inflate the parental risk figures.

The parenting corpus is small (139 posts) compared to the teenager corpus (1,489 posts), so figures on the parenting side are more noisy. This number of posts is despite retaining every relevant post from six parenting communities over three and a half years, drawn from 386,552 posts. So this scarcity is itself a result, indicating that youth AI companion use is discussed far less often in parenting communities than in teenager ones.

Seven of our 20 codes appear in fewer than five posts in the 100-post validation set, where $\kappa$ is too unstable to interpret; we report these as illustrative and draw no quantitative conclusions from them. Among the remainder, human agreement is borderline for harmful content and advice ($\kappa = 0.48$), and LLM-human agreement is moderate for social displacement ($\kappa = 0.58$), emotional support and companionship as a benefit ($\kappa = 0.57$), and entertainment and creative outlet ($\kappa = 0.59$).

Our work also carries several potential negative societal impacts. First, our finding that parents are largely unaware of certain aspects of adolescent use could be read as a case for more invasive monitoring, whereas we argue for the opposite: more open conversation between parents and children about AI companions. Relatedly, our findings should not be taken as an argument for outright restriction, since AI companions can serve as a valuable source of support for adolescents who lack other options. Third, the benefits we surface could be selectively cited to downplay the risks. Finally, our filtering method and codebook could be misused to identify and target vulnerable adolescents at scale from public posts---by monitoring software vendors, platforms, or anyone seeking to profile minors.

\section*{Acknowledgments}
This work was supported by the National Science Foundation (IIS-2229881), the National Institutes of Health (R35GM146974), a Moderna Fellowship Award, and a German National Academy of Sciences Leopoldina postdoctoral fellowship grant (LPDS 2024-06).

\bibliography{aaai2026}

\appendix

\section{Posts Dropped at Each Filtering Stage}
\label{app:attrition}
\begin{table*}[t]
\centering
\footnotesize
\setlength{\tabcolsep}{4pt}
\begin{tabular}{@{}lrrrrrrrr@{}}
\toprule
& & \multicolumn{6}{c}{Dropped by stage (\% of input)} & \\
\cmidrule(lr){3-8}
Subreddit & Posts & Cross-post & Removed & Bot/mod & Link/image & Duplicate & Short & Retained (\%) \\
\midrule
\multicolumn{9}{l}{\emph{Teenager subreddits}} \\
r/teenagers & 2,452,475 & $<$0.1 & 23.7 & $<$0.1 & 20.3 & 0.9 & 9.5 & 1,119,102 (45.6) \\
r/TeenagersButBetter & 212,546 & 5.6 & 20.7 & $<$0.1 & 51.3 & 0.2 & 2.1 & 42,682 (20.1) \\
\addlinespace
\multicolumn{9}{l}{\emph{Parenting subreddits}} \\
r/Parenting & 274,867 & $<$0.1 & 32.9 & 0.1 & 0.0 & 0.3 & $<$0.1 & 183,389 (66.7) \\
r/Mommit & 136,662 & $<$0.1 & 22.6 & 0.2 & 0.8 & 0.3 & 0.1 & 103,910 (76.0) \\
r/daddit & 121,232 & $<$0.1 & 13.9 & 0.1 & 18.4 & 0.3 & 0.1 & 81,417 (67.2) \\
r/Parents & 22,381 & 7.8 & 23.2 & 0.1 & 9.5 & 0.5 & 0.4 & 13,076 (58.4) \\
r/parentingteenagers & 8,288 & $<$0.1 & 73.4 & $<$0.1 & 0.0 & 0.1 & $<$0.1 & 2,199 (26.5) \\
r/raisingkids & 5,557 & 14.6 & 25.7 & 4.4 & 8.1 & 0.8 & 0.5 & 2,561 (46.1) \\
\midrule
\textbf{All} & 3,234,008 & 0.5 & 24.0 & $<$0.1 & 19.6 & 0.7 & 7.4 & \textbf{1,548,336 (47.9)} \\
\bottomrule
\end{tabular}
\caption{Number of posts dropped at each filtering stage.}
\label{tab:attrition}
\end{table*}

\section{Posts Matching Each Term}
\begin{table}[t]
\centering
\footnotesize
\setlength{\tabcolsep}{4pt}
\begin{tabular}{@{}l >{\raggedright\arraybackslash}p{0.66\columnwidth}@{}}
\toprule
\multicolumn{2}{@{}l@{}}{\emph{AI platforms}} \\
\addlinespace[2pt]
Character.AI    & \texttt{character.ai}, \texttt{character ai}, \texttt{characterai}, \texttt{c.ai}, \texttt{c ai}, \texttt{cai}, \texttt{char.ai}, \texttt{charai}, \texttt{char ai} \\
Replika         & \texttt{replika}, \texttt{replikaai}, \texttt{replika.ai} \\
Snapchat My AI  & \texttt{my.ai}, \texttt{myai}, \texttt{snapchat.ai}, \texttt{snapchat ai}, \texttt{snapchatai}, \texttt{snap.ai}, \texttt{snap ai}, \texttt{snapai} \\
Meta AI         & \texttt{meta.ai}, \texttt{meta ai}, \texttt{metaai} \\
Talkie          & \texttt{talkie ai} \\
Chai            & \texttt{chai ai}, \texttt{chai app}, \texttt{chai bot} \\
Janitor AI      & \texttt{janitor.ai}, \texttt{janitor ai}, \texttt{janitorai} \\
PolyBuzz        & \texttt{polybuzz}, \texttt{polybuzz.ai}, \texttt{poly.ai}, \texttt{poly ai}, \texttt{polyai} \\
Nomi AI         & \texttt{nomi.ai}, \texttt{nomi ai}, \texttt{nomiai} \\
Kindroid        & \texttt{kindroid} \\
SpicyChat       & \texttt{spicy.chat}, \texttt{spicy chat}, \texttt{spicychat} \\
Candy AI        & \texttt{candy.ai}, \texttt{candy ai}, \texttt{candyai} \\
ChatGPT         & \texttt{chatgpt}, \texttt{chat gpt} \\
Gemini          & \texttt{gemini} \\
Claude          & \texttt{claude} \\
Grok            & \texttt{grok} \\
\midrule
\multicolumn{2}{@{}l@{}}{\emph{Generic AI terms}} \\
\addlinespace[2pt]
AI              & \texttt{ai} \\
chatbot(s)      & \texttt{chatbot}, \texttt{chatbots} \\
bot(s)          & \texttt{bot}, \texttt{bots} \\
\bottomrule
\end{tabular}
\caption{Terms used to select posts for analysis.}
\label{tab:keywords}
\end{table}

\label{app:posts-matching}
\begin{table*}[t]
\centering
\footnotesize
\setlength{\tabcolsep}{4pt}
\begin{tabular}{@{}lrrrrrrrrr@{}}
\toprule
Term & r/teenagers & \shortstack[r]{r/Teenagers\\ButBetter} & r/Parenting & r/Mommit & r/daddit & r/Parents & r/raisingkids & \shortstack[r]{r/parenting\\teenagers} & \textbf{All} \\
\midrule
\emph{Posts} & 1,119,102 & 42,682 & 183,389 & 103,910 & 81,417 & 13,076 & 2,561 & 2,199 & \textbf{1,548,336} \\
\midrule
\emph{Dedicated companion apps} & 615 & 56 & 23 & 3 & 1 & 5 & 1 & 0 & \textbf{704} \\
\quad Character.AI & 570 & 52 & 18 & 1 & 1 & 5 & 1 & 0 & \textbf{648} \\
\quad Replika & 5 & 3 & 0 & 0 & 0 & 0 & 0 & 0 & \textbf{8} \\
\quad Talkie & 1 & 0 & 0 & 0 & 0 & 0 & 0 & 0 & \textbf{1} \\
\quad Chai & 9 & 0 & 0 & 0 & 0 & 0 & 0 & 0 & \textbf{9} \\
\quad Janitor AI & 22 & 4 & 1 & 0 & 0 & 0 & 0 & 0 & \textbf{27} \\
\quad PolyBuzz & 19 & 1 & 5 & 2 & 0 & 0 & 0 & 0 & \textbf{27} \\
\quad Nomi AI & 0 & 0 & 0 & 0 & 0 & 0 & 0 & 0 & \textbf{0} \\
\quad Kindroid & 0 & 0 & 0 & 0 & 0 & 0 & 0 & 0 & \textbf{0} \\
\quad SpicyChat & 1 & 0 & 1 & 0 & 0 & 0 & 0 & 0 & \textbf{2} \\
\quad Candy AI & 2 & 0 & 0 & 0 & 0 & 0 & 0 & 0 & \textbf{2} \\
\addlinespace
\emph{General-purpose assistants} & 1,612 & 139 & 148 & 108 & 202 & 35 & 20 & 5 & \textbf{2,269} \\
\quad ChatGPT & 1,386 & 123 & 131 & 98 & 171 & 34 & 17 & 5 & \textbf{1,965} \\
\quad Gemini & 111 & 7 & 15 & 9 & 19 & 4 & 3 & 0 & \textbf{168} \\
\quad Claude & 45 & 2 & 3 & 0 & 9 & 2 & 2 & 0 & \textbf{63} \\
\quad Grok & 41 & 7 & 0 & 0 & 5 & 0 & 0 & 0 & \textbf{53} \\
\quad Snapchat My AI & 63 & 1 & 1 & 0 & 1 & 0 & 0 & 0 & \textbf{66} \\
\quad Meta AI & 9 & 0 & 1 & 1 & 0 & 1 & 0 & 0 & \textbf{12} \\
\addlinespace
\emph{Generic AI terms} & 6,876 & 603 & 724 & 255 & 429 & 210 & 76 & 21 & \textbf{9,194} \\
\quad chatbot(s) & 196 & 23 & 26 & 5 & 9 & 10 & 8 & 3 & \textbf{280} \\
\quad bot(s) & 1,865 & 115 & 86 & 33 & 70 & 8 & 0 & 2 & \textbf{2,179} \\
\quad AI & 5,195 & 518 & 654 & 228 & 367 & 203 & 76 & 20 & \textbf{7,261} \\
\midrule
\textbf{Any term} & \textbf{8,027} & \textbf{698} & \textbf{814} & \textbf{338} & \textbf{575} & \textbf{223} & \textbf{83} & \textbf{22} & \textbf{10,780} \\
\bottomrule
\end{tabular}
\caption{Posts matching each term, by subreddit. Counts are non-exclusive; a post matching several terms appears in each corresponding row. Dedicated companion apps are platforms whose primary purpose is companionship or social, emotional, or sexual chat; general-purpose assistants are included because such tools are often used as companions.}\label{tab:termcounts-full}
\end{table*}

\section{Prompts}
\label{app:prompts}

\subsection{Relevance Filtering Prompt}
\label{app:filter-prompt}
\begin{quote}
\small
Does this Reddit post mention an AI system/tool used as an AI companion or for social, emotional, or therapeutic purposes?

Answer TRUE if EITHER applies:
\begin{itemize}
  \item MENTION: The post names or clearly refers to a product
    whose primary purpose is AI companionship or social/emotional/sexual chat---e.g., Character.AI (c.ai), Replika, Snapchat My AI, Talkie, Chai,
    Janitor AI, PolyBuzz/Poly.AI, Nomi, Kindroid, Spicychat, Candy AI. A bare
    mention counts, even if the post is about the app being down, the price, a
    bug, a ban, or an offhand reference in a list of apps. No description of
    use is required.
  \item DESCRIBED USE: The post describes someone using any AI
    system (including general assistants like ChatGPT, Gemini, Claude, Grok,
    or Meta AI) for companionship, venting, emotional support, sexual
    roleplay, or therapy.
\end{itemize}

Answer FALSE if none of that applies. In particular, FALSE if:
\begin{itemize}
  \item The only AI present is a general-purpose tool used for non-social tasks (homework, code, image generation, search, drafting messages, storytelling, content generation).
  \item There is no AI at all.
  \item The post advertises, pitches, or solicits feedback on an app the author is building, and no real person's actual use is described.
\end{itemize}

EXCLUSION---apply only to (b), never to (a): if the described use is by a parent and no one else is described as using it, answer FALSE. This holds however emotional the use is---a parent who calls AI their therapist, best friend, or substitute for friends is still FALSE. Check this last, on every post you were about to call TRUE under (b).

Respond with ONLY true or false, then one short sentence of
justification.
\end{quote}

\subsection{Codebook Application Prompt}
\label{app:coding-prompt}
\begin{quote}
\small
You are coding Reddit posts for a study comparing how teenagers and parents talk about AI companions (Character.AI, Replika, Snapchat My AI, ChatGPT, etc.).

For each category below, apply every code whose definition fits the post. A post may take several codes in a category, or none at all---return an empty list when nothing fits. Quote the words in the post that support each code. The examples
under each definition are illustrations, not an exhaustive list.

Every code describes an AI system, and only the parts of the post that involve an AI can be evidence for one. Activity with no AI in it---roleplay or games with friends, time on other apps, talking to a person---never earns a code, however well it matches the wording of a definition. The quote you give must itself be about the AI; if you cannot quote such a passage, do not apply the code.

\medskip
\textbf{Modes of use}\\
What mode of use is being discussed by the author of this post?

\begin{itemize}
  \item \textbf{Emotional support and personal advice:} Turning to the AI to cope with distress, e.g.\ venting, comfort, loneliness, anxiety, therapy-like use; or asking the AI for guidance on a real-world or interpersonal decision. Exclude: homework, task help, information lookup.
    \begin{itemize}
      \item ``Been using c.ai since 10th grade as cure to my loneliness''
      \item ``My teenager uses AI to deal with school stress.''
      \item ``I personally use it to talk about things like love, gaming or
        religion and stuff. And to be honest, it gives really good advice''
      \item ``I asked chat gpt if i should text the guy or if im being crazy''
    \end{itemize}
  \item \textbf{Roleplay and imaginative scenarios:} Fictional or character-driven play with the AI: scenarios, personas, in-character interaction, including creative outlet. Does not include sexual roleplay.
    \begin{itemize}
      \item ``the only reason i do c.ai is because I have no friends to roleplay.''
      \item ``She uses it as an outlet for her creativity.''
    \end{itemize}
  \item \textbf{Friend and best friend:} The AI is explicitly named as a friend or companion.
    \begin{itemize}
      \item ``Chatgpt has been my friend.''
      \item ``He said Alexa was his best friend.''
    \end{itemize}
  \item \textbf{Romantic and sexual interactions:} Romantic or sexual interaction with the AI: partner, girlfriend/boyfriend, flirting, erotic chat.
    \begin{itemize}
      \item ``I've never had a girlfriend in my life and CAI makes me feel better,
        like I'm loved.''
      \item ``She was asking Snapchat AI about sex.''
    \end{itemize}
\end{itemize}

\medskip
\textbf{Risks}\\
What risk posed by AI is being perceived by the author of this post?

\begin{itemize}
  \item \textbf{Data harvesting and privacy:} Concern about collection of data during interaction with AI and it being turned into targeted marketing, becoming public after a security breach, etc.
    \begin{itemize}
      \item ``Yeah, ChatGPT steals your data and trains models with it.''
      \item ``The app has access to her location, browsing history, everything.''
    \end{itemize}
  \item \textbf{Sexual and age-inappropriate interactions:} Concern about grooming, seduction, or coercion into sexual acts. Warping teens' ideas about healthy sexual behavior. Exposure to content that is inappropriate for the child's age (sexual, violent, or otherwise disturbing).
    \begin{itemize}
      \item ``AI generated images of women aren't any less harmful in regard to
        warping expectations of what real bodies will look like.''
      \item ``She was having inappropriate sexual conversations with AI.''
    \end{itemize}
  \item \textbf{Attachment and dependence:} Concern about AI exploiting natural human tendency for attachment for the purpose of generating engagement, and concern with compulsive use (how much, how often, and being unable to stop), driven by always-supportive, always-available AI with features that promote engagement. Includes posts mentioning a specific amount of time using AI and framing that amount as excessive.
    \begin{itemize}
      \item ``You are going to heavily risk developing psychosis or symptoms
        similar to it if you develop an attachment to ai.''
      \item ``My son is becoming too attached to our Alexa.''
      \item ``I've been addicted to ai for so long.''
      \item ``My sister uses Character ai for 3 hours every night.''
    \end{itemize}
  \item \textbf{Harmful content and advice:} Concern that the AI encourages or normalizes harm to self or others (e.g., self-harm, suicide, disordered eating, violence).
    \begin{itemize}
      \item ``It was giving her information on how to hide an eating disorder.''
      \item ``I recently read about the teenager who took his own life after using
        chatGPT for emotional support.''
    \end{itemize}
  \item \textbf{Sycophancy:} Concern that the AI always agrees and never pushes back, whatever the user already thinks.
    \begin{itemize}
      \item ``Sure AI will listen and not criticize, but what if it reinforces your
        negative thoughts?''
      \item ``It never pushes back or tells them they're wrong.''
    \end{itemize}
  \item \textbf{Social displacement:} Concern that time with the AI is substituting for, and reducing, contact with real people.
    \begin{itemize}
      \item ``Why are we making it worse by having people talk to their ai
        companions instead of real people?''
      \item ``I'm worried that relying too much on ChatGPT might make him even more
        introverted.''
    \end{itemize}
\end{itemize}

\medskip
\textbf{Benefits}\\
What benefit of AI is being stated or clearly implied by the author of this post?

\begin{itemize}
  \item \textbf{Emotional support and companionship:} The AI provides comfort, a sense of being heard, or help managing distress; or fills an absence of company, giving the author someone to talk to.
    \begin{itemize}
      \item ``it's not always agreeing with me, but it listens''
      \item ``I use PolyBuzz to make myself feel needed''
      \item ``I downloaded those damn AI apps so I can talk to people again''
      \item ``I have no-one else to talk to''
    \end{itemize}
  \item \textbf{Useful advice:} The AI gives useful guidance on a real-world or interpersonal decision.
    \begin{itemize}
      \item ``I use it to talk about things like love, gaming or religion\ldots{} it gives really good advice''
      \item ``like an imaginary friend, except this one can actually provide useful information''
    \end{itemize}
  \item \textbf{Entertainment and creative outlet:} The AI is fun, passes the time, or keeps a child occupied; or is a valuable outlet for the user's imagination and creativity.
    \begin{itemize}
      \item ``It kept him entertained the whole time I was dealing with the dishes''
      \item ``Im so bored i started talking to Ai''
      \item ``It's creative, she has an outlet''
      \item ``having playful creative ideas''
    \end{itemize}
  \item \textbf{Low-stakes and judgment-free interaction:} The AI is valuable because interacting with it carries no social pressure, embarrassment or fear of judgment.
    \begin{itemize}
      \item ``I could share my ideas with it because I'm very shy to show them to other people''
      \item ``a patient, non-judgmental space to explore topics''
    \end{itemize}
  \item \textbf{Social skills practice:} The AI is valuable to rehearse social skills and build capacity for interaction with people.
    \begin{itemize}
      \item ``It's nice for people who are a bit less social or more introverted to practice some interactions and have some social skills''
      \item ``we've been working hard with him on responding to people when they talk and general ability to carry a conversation''
    \end{itemize}
  \item \textbf{Safer outlet and harm reduction:} The AI is framed as the lesser risk compared with the human alternative.
    \begin{itemize}
      \item ``TBH I'm less worried by the sexting AI, I'd rather that than the alternative''
      \item ``role playing sex helps me stop doing it''
    \end{itemize}
\end{itemize}

\medskip
\textbf{Mediation strategies}\\
What parental mediation strategy is mentioned in this post?

\begin{itemize}
  \item \textbf{Instructive:} Parent actively discusses guidelines for AI companion usage, such as discussing usage rules or having a broader conversation about AI companion risks and features.
    \begin{itemize}
      \item ``I try to communicate with my child about this rather than block everything.''
      \item ``We've told her AI is dangerous.''
      \item ``I am unsure how to talk with her about it''
      \item ``We need to talk about boundaries with chat bots''
    \end{itemize}
  \item \textbf{Restrictive:} Parent limits child's access to AI companion, including requiring explicit parental permission before use.
    \begin{itemize}
      \item ``I have age control limits applied.''
      \item ``Should we block ChatGPT?''
    \end{itemize}
  \item \textbf{Co-use:} Joint interactions, such as parent and child using AI companion together or sharing its outputs with each other. The parent takes part in the interaction. They are doing it together, not just checking it. A parent who watches the child's session in order to catch problems is supervision, not co-use.
    \begin{itemize}
      \item ``After that deepfake incident, I sat with my son and we started watching fake videos together.''
    \end{itemize}
  \item \textbf{Supervision:} Parent monitors child's interactions with AI companion, screening prompts and chats, or only allowing child access to AI companion on shared device. Checking the phone, browser history, account, chat log, or alerts from monitoring software count. Listening in on or watching an ongoing session also counts.
    \begin{itemize}
      \item ``We use the `family computer' so it can be supervised.''
      \item ``I went over his phone''
      \item ``I have been listening to it and keeping an eye on it''
      \item ``Please monitor what apps your kids are using''
    \end{itemize}
\end{itemize}
\end{quote}

\section{Coding Instructions}
\label{app:coding-instructions}
\begin{quote}
\small
This is a hybrid (mix of deductive and inductive) coding session, so for each category (except Benefits) we are starting from a codebook from the literature, and we will extend those based on what we find in the data.

\paragraph{Coding process.}
\begin{itemize}
  \item There are 4 categories, reflecting the research questions: Modes of use, Risks, Benefits, and Mediation strategies.
  \item Each post can have zero, one, or multiple codes per category.
  \item Use existing codes if they fit, or write new codes if they don't.
  \item Select an evidence span in the text for each code.
\end{itemize}

\paragraph{General guidelines.}
\begin{itemize}
  \item Skip posts that are only about an adult's usage of AI.
  \item Code posts about sibling's / someone else's / general use of AI.
  \item Code even if in the past, e.g.\ ``I used to have an AI girlfriend but now I stopped'' counts.
\end{itemize}

\paragraph{Per-category guidelines.}
\begin{itemize}
  \item \textbf{Modes of use:}
    \begin{itemize}
      \item Answer the question: What mode of use is being discussed by the author of this post?
    \end{itemize}

  \item \textbf{Risks:}
    \begin{itemize}
      \item Answer the question: What risk posed by AI is being perceived by the author of this post?
      \item This does not include risky behavior which is discussed/exhibited but which the author is unaware of. E.g.: ``ChatGPT does all my homework'', ``I love my AI girlfriend'' by themselves do not qualify.
    \end{itemize}

  \item \textbf{Benefits:}
    \begin{itemize}
      \item Answer the question: What benefit of AI is being stated or clearly implied by the author of this post?
      \item Code even if the overall valence of a post is negative.
    \end{itemize}

  \item \textbf{Mediation strategies:}
    \begin{itemize}
      \item Answer the question: What parental mediation strategy is mentioned in this post?
      \item Code mediation from sibling / young relative.
      \item Don't code mediation from therapist.
      \item Bringing child to therapist does not qualify as a form of mediation.
      \item Code any mediation strategy that is somehow connected to AI companion use, e.g.:
        \begin{itemize}
          \item Code ``I found out he was using AI because I checked his phone'' as supervision.
          \item Don't code ``banning from using TikTok'' as restrictive.
        \end{itemize}
    \end{itemize}
\end{itemize}
\end{quote}

\section{Full Codebook}
\label{app:codebook}
Tables~\ref{tab:cb-modes}--\ref{tab:cb-mediation} give the full definition and an example for each of the 20 codes summarized in Table~\ref{tab:codes}.

\begin{table*}[t]
\centering
\small
\begin{tabularx}{\textwidth}{@{}p{2.6cm} >{\raggedright\arraybackslash}X >{\raggedright\arraybackslash}X@{}}
\toprule
Code & Definition & Example \\
\midrule
Emotional support and personal advice &
Turning to the AI to cope with distress, e.g.\ venting, comfort, loneliness, anxiety, therapy-like use; or asking the AI for guidance on a real-world or interpersonal decision. Excludes homework, task help, and information lookup. &
``My son uses AI to cope with school stress.'' \\
\addlinespace
Roleplay and imaginative &
Fictional or character-driven play with the AI: scenarios, personas, in-character interaction, including creative outlet. Does not include sexual roleplay. &
``She uses CharacterAI as an outlet for her creativity.'' \\
\addlinespace
Friend and best friend &
The AI is explicitly named as a friend or companion. &
``I see Chatgpt as a friend.'' \\
\addlinespace
Romantic and sexual interactions &
Romantic or sexual interaction with the AI: partner, girlfriend/boyfriend, flirting, sexual chat. &
``I found him asking Meta AI about sex.'' \\
\bottomrule
\end{tabularx}
\caption{Codebook for modes of use.}
\label{tab:cb-modes}
\end{table*}

\begin{table*}[t]
\centering
\small
\begin{tabularx}{\textwidth}{@{}p{2.6cm} >{\raggedright\arraybackslash}X >{\raggedright\arraybackslash}X@{}}
\toprule
Code & Definition & Example \\
\midrule
Data harvesting and privacy &
Concern about collection of data during interaction with AI and it being turned into targeted marketing, becoming public after a security breach, etc. &
``Claude steals your personal data and improves its models with it.'' \\
\addlinespace
Sexual and age-inappropriate interactions &
Concern about grooming, seduction, or coercion into sexual acts; warping teens' ideas about healthy sexual behavior; exposure to content inappropriate for the child's age (sexual, violent, or otherwise disturbing). &
``AI generated images warp expectations of what real bodies look like.'' \\
\addlinespace
Attachment and dependence &
Concern about AI exploiting the human tendency for attachment in order to generate engagement, and concern with compulsive use (how much, how often, being unable to stop). Includes posts that mention a specific amount of use and frame it as excessive. &
``I am completely addicted to ai.'' \\
\addlinespace
Harmful content and advice &
Concern that the AI encourages or normalizes harm to self or others (e.g., self-harm, suicide, violence). &
``The AI told her how to hide self-harm.'' \\
\addlinespace
Sycophancy &
Concern that the AI always agrees and never pushes back, whatever the user already thinks. &
``AI never pushes back or tells you you're wrong.'' \\
\addlinespace
Social displacement &
Concern that time with the AI is substituting for, and reducing, contact with real people. &
``Using ChatGPT so much might make him even more introverted.'' \\
\bottomrule
\end{tabularx}
\caption{Codebook for perceived risks.}
\label{tab:cb-risks}
\end{table*}

\begin{table*}[t]
\centering
\small
\begin{tabularx}{\textwidth}{@{}p{2.6cm} >{\raggedright\arraybackslash}X >{\raggedright\arraybackslash}X@{}}
\toprule
Code & Definition & Example \\
\midrule
Emotional support and companionship &
The AI provides comfort, a sense of being heard, or help managing distress; or fills an absence of company, giving the author someone to talk to. &
``It does not always agree with me, but it listens.'' \\
\addlinespace
Useful advice &
The AI gives useful guidance on a real-world or interpersonal question. &
``I talk with AI about things like love and religion\ldots{} I got good advice.'' \\
\addlinespace
Entertainment and creative outlet &
The AI is fun, passes the time, or keeps a child occupied; or is a valuable outlet for the user's imagination and creativity. &
``It gives her a creative outlet.'' \\
\addlinespace
Low-stakes, judgment-free interaction &
The AI is valuable because interacting with it carries no social pressure, embarrassment, or fear of judgement. &
``It's a patient, non-judgmental companion to explore with.'' \\
\addlinespace
Social skills practice &
The AI is valuable to rehearse social skills and build capacity for interaction with people. &
``It's useful for less social people to practice interactions.'' \\
\addlinespace
Safer outlet and harm reduction &
The AI is framed as the lesser risk compared with the human alternative. &
``I prefer her sexting AI than the alternative.'' \\
\bottomrule
\end{tabularx}
\caption{Codebook for perceived benefits.}
\label{tab:cb-benefits}
\end{table*}

\begin{table*}[t]
\centering
\small
\begin{tabularx}{\textwidth}{@{}p{2.6cm} >{\raggedright\arraybackslash}X >{\raggedright\arraybackslash}X@{}}
\toprule
Code & Definition & Examples \\
\midrule
Instructive &
Parent actively discusses guidelines for AI companion usage, such as discussing usage rules or having a broader conversation about AI companion risks. &
``I try to talk to my kid about it rather than block everything.'' \\
\addlinespace
Restrictive &
Parent limits the child's access to the AI, including requiring explicit parental permission before use. &
``I use age control limits.'' \\
\addlinespace
Co-use &
Joint interactions: parent and child use the AI companion together or share its outputs with each other. The parent takes part in the interaction rather than only checking it; watching a session in order to catch problems is supervision, not co-use. &
``After finding the deepfakes, I sat with my daughter and we watched fake videos together.'' \\
\addlinespace
Supervision &
Parent monitors the child's interactions with the AI companion: screening prompts and chats, checking the phone, browser history, account, or chat log, receiving alerts from monitoring software, listening in on or watching an ongoing session, or only allowing access on a shared device. &
``I try to keep an eye on the app.'' \\
\bottomrule
\end{tabularx}
\caption{Codebook for parental mediation strategies.}
\label{tab:cb-mediation}
\end{table*}

\section{Codebook Development}
\label{app:codebook-development}
\begin{table*}[t]
\centering
\small
\begin{tabularx}{\textwidth}{@{}llX@{}}
\toprule
Initial code & Action & Final code and rationale \\
\midrule
\multicolumn{3}{@{}l}{\emph{Modes of use (RQ1)}} \\
\addlinespace[2pt]
Conversation and social practice & Dropped & --- \newline
Too generic; nearly all posts would qualify. \\
Emotional and mental-health support & Broadened & Emotional support and personal advice \newline
Extended to cover advice seeking. \\
Roleplay and imaginative scenarios & Redefined & Roleplay and imaginative \newline
Extended to cover creative writing; sexual roleplay explicitly excluded to avoid overlap with \emph{Romantic and sexual interactions}. \\
Friend and best friend & Retained & Friend and best friend \\
Romantic and sexual interactions & Retained & Romantic and sexual interactions \\
\addlinespace
\multicolumn{3}{@{}l}{\emph{Risks (RQ2)}} \\
\addlinespace[2pt]
Data harvesting and privacy & Retained & Data harvesting and privacy \\
Overtrust & Dropped & --- \newline
Too generic; overlapped with \emph{Attachment and dependence} and \emph{Harmful content and advice}. \\
Sexual interactions & Broadened & Sexual and age-inappropriate interactions \newline
Some posts described age-inappropriate but non-sexual content, such as violence. \\
Attachment and manipulation \\
Dependence and compulsive use & Split and merged & Attachment and dependence \newline
Manipulation separated out into \emph{Harmful content and advice}; the remainder merged with \emph{Dependence and compulsive use}, which was closely related. \\
--- & Added & Harmful content and advice \newline
Posts describing a system encouraging harmful action fit no existing code; subsumes manipulation. \\
--- & Added & Sycophancy \newline
Recurred in parenting subreddits. \\
--- & Added & Social displacement \newline
Loneliness was a recurrent theme in teenager subreddits. \\
\addlinespace
\multicolumn{3}{@{}l}{\emph{Benefits (RQ2)}} \\
\addlinespace[2pt]
--- & Added & Emotional support and companionship \newline
Frequent in teenager subreddits. \\
--- & Added & Useful advice \newline
Occasional in teenager subreddits. \\
--- & Added & Entertainment and creative outlet \newline
Frequent in both community types. \\
--- & Added & Low-stakes, judgment-free interaction \newline
Occasional in parenting subreddits. \\
--- & Added & Social skills practice \newline
Occasional in parenting subreddits. \\
--- & Added & Safer outlet and harm reduction \newline
Occasional in parenting subreddits. \\
\addlinespace
\multicolumn{3}{@{}l}{\emph{Mediation strategies (RQ3)}} \\
\addlinespace[2pt]
Instructive & Retained & Instructive \\
Restrictive & Retained & Restrictive \\
Co-use & Narrowed & Co-use \newline
Restricted to cases where the parent participates in the interaction, to avoid overlap with \emph{Supervision}. \\
Supervision & Narrowed & Supervision \newline
Restricted to cases where the parent does not participate in the interaction, to avoid overlap with \emph{Co-use}. \\
\bottomrule
\end{tabularx}
\caption{Audit trail of codebook development.}
\label{tab:audit-trail}
\end{table*}

\end{document}